# Probing intracellular mass density fluctuation through confocal microscopy: application in cancer diagnostics as a case study

Peeyush Sahay[1], Aditya Ganju[2], Hemendra M. Ghimire[1], Huda Almabadi[1],
Murali M. Yallappu[2], Omar Skalli[3], Meena Jaggi[2], Subhash C. Chauhan[2], Prabhakar Pradhan[1*]

[1]Department of Physics, BioNanoPhotonics Laboratory, University of Memphis, Memphis, TN 38152, USA
[2]Department of Pharmaceutical Sciences and the Center for Cancer Research, College of Pharmacy, University of Tennessee Health Science Center, Memphis, TN 38163, USA
[3]Department of Life Sciences and Integrated Microscopy Center, University of Memphis, Memphis, TN 38152, USA
*Corresponding author: ppradhan@memphis.edu

## Abstract

Intracellular structural alterations are hallmark of several disease conditions and treatment modalities. However, robust methods to quantify these changes are scarce. In view of this, we introduce a new method to quantify structural alterations in biological cells through the widely used confocal microscopy. This novel method employs optical eigenfunctions localization properties of cells and quantifies the degree of structural alterations, in terms of nano- to micron scale intracellular mass density fluctuations, in *one single parameter.* Such approach allows a powerful way to compare changing structures in heterogeneous cellular media irrespective of the origin of the cause. As a case study, we demonstrate its applicability in cancer detection with breast and prostate cancer cases of different tumorigenicity levels. Adding new dimensions to the confocal based studies, this technique has potentially significant applications in areas ranging from disease diagnostics to therapeutic studies, such as patient prognosis and assessing survival potential.

# INTRODUCTION

Structural alterations in cellular organelles are commonly occurring phenomena. Events ranging from exposure of radiation and heat, disease processes, such as cancer and inflammation, as well as response to therapeutic treatments, etc., are known to proceed with such alterations affecting the cell's dynamics and function[1–6]. These changes are usually characterized qualitatively by light microscopy but methods to characterize them quantitatively are scarce. Although, recently there have been few attempts to quantify the degree of intracellular structural alterations[7–9], a robust method capable of providing an easy comparison of changing structures in cell like media is still missing. The complexity involved in quantifying structural alterations, also termed as structural disorder, arises due to the heterogeneous nature of the cell, which requires the use of various spatial mass density correlations, such as exponential decay, stretched exponential decay, morphological multi-fractality, etc[10,11]. Such situation demands multiple length scales to be considered for characterizing the system. This calls for the development of a method which not only efficiently quantifies the degree of intracellular structural disorder by taking into account all the structural statistics of a cell, but also provides an easy comparison technique, for example, by measuring in a single parameter. Additionally, the method should also be practical and compatible with common imaging modalities used in biomedical research.

With these considerations in mind, we have developed a novel method to quantify structural disorder inside cells through the widely used confocal microscopy. The method employs analyzing the optical eigenfunctions localization or light localization properties of a refractive index disordered optical lattice system constructed via confocal imaging of the cells. The quantification is performed in one single parameter, namely the average inverse participation ratio (IPR) value of the optical lattice system. The IPR value, calculated through the statistical analysis of the eigenfunctions of the lattice system, is a measure of the strength of light localization in an optical lattice, which in turn is an indicative of the degree of effective disorder in the system[12,13]. Such approach provides a better tool for comparing changing structure of a heterogeneous cellular media irrespective of its origin of cause. Therefore, this method is quite versatile and universal, potentially capable of applying for a spectrum of research and diagnostic purposes, where the cellular structural deformities occurred irrespective of the cause.

Analysis of disordered systems through its optical eigenfunction localization or light localization properties is an extensively used technique in condensed matter physics[14], and herein we for the first time employ such powerful approach to characterize structural changes in biological systems through the commonly used confocal microscopy. In this article, we first briefly describe the background, theoretical framework, and working principle of the technique along with a discussion of its importance. We then present a case study of its application in cancer detection. We demonstrate that this technique can be efficiently employed to quantify nano- to micron scale levels of spatial mass density fluctuations, associated with the resolution of confocal microscopy, in nuclear DNA in cells at different stages of carcinogenesis. Thus, potentially presenting a novel method for cancer detection. Finally, we have also discussed the potential applications of the technique to a wide range of research and diagnostic areas such as cancer detection, radiation damage of cells, cells in tress, sickle cell anemia etc., based on cell/tissue structural alteration/deformities, where confocal microscopy is used as standard characterization modality.

**METHODs**

*Confocal microscopy imaging as tool to quantify structural disorder in cells and tissues*: Confocal microscopy (CFM) is one of the widely used fluorescence microscopic technique for studying biological samples. The pinhole system used in CFM allows rejecting the out of focus light, which results in higher resolution, better SNR, and therefore relatively accurate quantitative measurements[15,16]. To perform quantitative measurements through CFM the underlying idea is that the fluorophores molecules that is used for staining the target molecules stoichiometrically binds with them; the fluorescence intensity detected in the image plane correlates with the fluorophores concentration, therefore, pixel-wise intensity analysis of a 2D confocal image is representative of local concentration or mass density distribution inside the cell[15–21]. The pixel intensity detected in the confocal image plane, $I_{det}$, depends upon the illumination intensity in the sample and the fluorescence intensity from the voxel at (x,y) at a depth z inside the sample [22]. In the present work, we make approximations that (i) Average intensity fluorescing out of a voxel is linearly proportional to the mass-density (local concentration) of the target molecules inside the voxel, (ii) Fluorophores are not photo bleached or saturated (iii) Illumination intensity and SNR remains uniform while scanning one horizontal plane in the sample (i.e., varying voxel position in x-y, z=fixed). Under such considerations, the pixel intensity measured in confocal imaging can be proportionally equated with the mass-density of the stained molecules inside the voxel emitting the fluorescent signal; and the fluctuations in the recorded intensity pattern of a 2d, x-y cell plane can be attributed to the voxel-to-voxel mass-density variation inside the sample. Mathematically, $dI_{CFM}(r)/<I>_{CFM} \propto d\rho_{cell}(r)/<\rho>_{cell}$; where $<I>_{CFM}$ and $d_{CFM}I(r)$ represent the average intensity of the confocal image and intensity fluctuations at positions $r$, respectively, while, $<\rho>_{cell}$ and $d\rho_{cell}(r)$ represent average mass density of the fluorophore molecules, thus that of the stained molecules as well, inside the sample and its fluctuations at positions $r$, respectively.

*Refractive index optical lattice and optical potential derivation from confocal imaging intensity data:* It has been shown that the refractive index, $n(r)$, inside a cell is linearly proportional to local mass-density, $\rho_{cell}(r)$, of macromolecules, e.g. DNA, RNA, lipids, etc., $n(r) = n_0 + dn(r) = <\rho>_{cell} + \alpha \, d\rho_{cell}(r)$ [23,24]; where $n_0$ is the average RI of the cell; $dn(r)$ represents RI fluctuation at the position $r$; and α is the proportionality constant. If fluctuation in RI is represented by $dn/n_0$, then we can write: $dn/n_0 \propto dI(r)_{CFM}/<I>_{CFM}$.

Figure 1 is a schematic representation to illustrate how confocal imaging is used to construct the refractive index optical lattice system. The z-stack confocal images are obtained by optical slicing of the sample (Fig. 1(a) and (b)). From the z-stack (Fig. 1(a)), 2-3 images are chosen from the middle of the stack. Figure 1(c) shows a representative confocal image of one cell nuclei. Figure 1(d) shows the intensity (or the refractive index) distribution in one particular row of the cell image in Figure 1(c). The terms $dn$ and $l_c$ represent the extent of fluctuation in intensity and its correlation length, respectively, along that row. As noted above, this intensity pattern depicts a linearly proportional mass density variation along that line (row) in the sample, and, in turn, depicts a proportional refractive index distribution along that row. Figure 1(e) shows a small cell area about 4 ×4 pixels in size of a confocal image (in 2D plan) from which an 4 × 4 two-dimensional (2D) optical grid is constructed by mapping point-to-point from the confocal image. The values of the points (pixels) in the constructed optical grid represent the proportional refractive index at those points, and are termed as 'optical potential'. Consequently, we obtain a refractive index matrix of the whole 2D sample. It is important to point out here that relative to 3D, light localization is stronger in 1D and 2D media[25]. In fact, light localization in weakly disordered media is a well-studied topic in soft condensed matter physics especially for 2D[12,26–28]. For a simple case with strong light localization properties, we therefore have considered analyzing the properties of 2D optical lattice system constructed from confocal imaging. In particular, we chose to analyze several 2D confocal images separately taken from the z-stacks, rather than considering the whole z-stack together, which forms the 3D shape of the cell

*Tight-binding model for the eigenfunctions of the optical/refractive index lattice/matrix and IPR calculations*: Having constructed the refractive index lattice, we employed the Anderson tight-binding model (TBM) Hamiltonian to analyze disorder in our system. The Maxwell light wave equation was solved in the constructed disordered optical lattice system from cell with closed boundary condition, and the eigenvalues and eigenfunctions of the system were obtained. Considering that $\varepsilon_i$ $(=dn_i/n_0)$ represents the onsite refractive index, or optical potential of the $i^{th}$ site, and that $|i\rangle$ represents the state of a photon at any arbitrary lattice site, the Hamiltonian for the entire lattice system can be written using the TBM when inter-lattice site hopping is restricted to only the nearest neighbor, given by,[12]

$$H = \sum_i \varepsilon_i |i\rangle\langle i| + t \sum_{\langle ij \rangle} |i\rangle\langle j| + |j\rangle\langle i| \qquad (1)$$

where $t$ is the inter-lattice site hopping or overlapping integral parameter (in Eq. (1)). From the above Hamiltonian, the average value of IPR is calculated [12,29],

$$<IPR(L)>_{\text{Pixel}} = \frac{1}{N}\sum_{i=1}^{N}\int_0^L\int_0^L E_i^4(x,y)\,dx\,dy \qquad (2)$$

where, in Eq. (2), $E_i$ is the $i^{th}$ eigenfunction of the lattice Hamiltonian, and $L \times L$ is the dimension of the sample. The IPR value of an eigenfunction provides the degree of localization of the eigenfunction, which, in turn, is a measure of the degree of disorder of the sample. Simply stated, a higher average IPR value suggests higher disorder in the system[12]. Typically, disorder inside a sample is characterized by $L_d$, which is a parameter of degree of structural disorder or "disorder strength" based on a Gaussian color noise disorder model with exponential decay correlation, $<dn(r)dn(r')> = <dn^2>exp(-|r-r'|/l_c)$, where $dn$ represents the fluctuation in refractive index, and $l_c$ represents the correlation length of that fluctuation. It can be shown that for 2D[14]

$$<IPR> \propto L_d \propto dn \times l_c \qquad (3)$$

*Cell culture, fluorescence staining, and confocal imaging:* All cell lines used in this paper were obtained from American Tissue and Cell Collection (ATCC). These included MCF 10A normal breast epithelial cells, MCF 7 metastatic breast carcinoma cells, PWR-1E normal prostate epithelial cells, LNCaP, DU145, and C4-2 metastatic prostate carcinoma cells. Cells were grown using standard cell culture procedures in media recommended by the ATCC. For microscopy, cells on a #1.5 cover glass were fixed in 4% paraformaldehyde in phosphate buffered saline (PBS) for 5 minutes, rinsed 3x5 minutes in PBS each and mounted in Prolong Diamond medium containing DAPI (4',6-diamidino-2-phenylindole), a fluorescent dye that strongly binds to nuclear DNA. Confocal fluorescence microscopy observations were carried out with a Nikon A1 confocal system equipped with a laser operating at 407 nm and a 63x, NA 1.4 oil objective corrected for red, blue, and violet chromatic aberrations.

**RESULT S AND DISCUSSION**

*Disorder analysis of breast control/cancer cells:* Confocal imaging of DAPI-stained nuclei in two breast epithelial cell lines, namely MCF-10A (control) and MCF-7 (cancer) were conducted. The horizontal plane resolution (x-y resolution) was ~0.2 × 0.2 (µm)$^2$, and the z-axis resolution was 0.13 µm. The pixel resolution of the images was 1024×1024. Several optical sections were obtained along the z-axis to capture the 3D structure of the nucleus. For every z-stack of a cell nucleus, 2-3 micrographs from the middle of the nucleus were used to perform the structural disorder analysis study on the x-y plane. Selection of the images from the middle of the stack was done to cover the maximum area, as well as changes in the nucleus. Using these conditions, structural disorder analysis was performed for a total of 10-12 cells for each of the cell lines, totaling > 25 2d confocal micrograph for each cell category. The results are shown in Figure 2.

Figure 2(a) and 2(b) are the 2D confocal images of DAPI-stained nuclei from MCF-10A and MCF-7cells, respectively, exhibiting heterogeneity in the staining intensity of nuclear DNA, indicating variations in concentration, i.e., mass density of DNA within the nucleus. Using the intensity values at each pixel, the structural disorder $L_d$ values i.e., mass density fluctuations, inside the nucleus of MCF-10A (Fig. 2(a)) and MCF-7 (Fig. 2(b)) cells were determined as described in Eq. 3. The corresponding $L_d$ images are shown in Figure 2(a') and (b'), respectively (measured for the sample size A= 0.4 × 0.4 (µm)$^2$). The areas colored in red in the $L_d$ images indicate regions with higher structural disorder, i.e., higher refractive index fluctuation region, and thus higher mass density fluctuations in that vicinity. The $L_d$ values calculated at different sample length scales $L$ (sample size $L \times L$ (µm)$^2$) for the two cell lines are plotted in Figure 2(c). The result shows that the structural disorder measured for MCF-7 cells at length scales ranging from 0.4 to 1.6 µm are higher than those of MCF-10A cells. Figure 2(d) represents the comparison bar graph between the $L_d$ values measured at the sample length 1.6 µm. The $L_d$ values calculated for MCF-7 cells (cancer) at the sample length 1.6 µm turned out to be significantly higher from those of MCF-10A cells (control). This reveals that the structural disorder that develops inside the cells during carcinogenesis increases with the progression of carcinogenesis, as it is evident with the MCF-7 and MCF-10A cells results, and that it can be quantified using confocal imaging. These results suggest that the method described herein has the potential to quantify the degree of tumorigenicity of cancer cells.

*Prostate normal and cancer cells:* Encouraged by the results of the breast cancer cell experiments, we subsequently use the technique to study human prostate normal and cancerous cells. However, unlike the study of breast normal/cancer cells imaged from one single nucleus, i.e., one single cell at a time, prostate cells were imaged in a matrix of cells on a slide in a single shot, with the image plane around the middle of the cells. The analyses were done by quantifying the structural disorder of all the cells together in the single image frame, as well as analyzing 8-12 single nuclei which were separately, but randomly, picked up from the grouping. This protocol reduced the extra effort of preparing and imaging one single cell at a time and provided a statistically better view by comparing the structural disorder in a distribution of similar cells in just one shot, portending, in turn, the quick and efficient analysis of structural disorder in cells and tissues. We used the well-characterized PWR normal prostate cells, in addition to tumorigenic and low metastatic LNCaP (AR-dependent), highly tumorigenic and highly metastatic DU145 (AR-independent), and highly tumorigenic and highly bone metastatic C4-2 (AR-independent) cell lines. These cancer cell lines are well-established *in vitro* models of early and advanced stage prostate cancer based on their cellular and molecular characteristics.

Figures 3(a)-(d) show the confocal images (micrograph) of cell matrix on glass slides. Figures 3(a), (b), (c), and (d) are the images of PWR (normal prostate cell), as well as the cancerous LNCaP, DU145, and C4-2 cell lines, respectively. The images shown in Figures 3(a'), (b'), (c'), and (d') are their corresponding $L_d$ images. As explained

above, the red-colored regions represent a higher level of spatial mass density fluctuation in that region. From the figures, it can be seen that the actual confocal images cannot be used for visible distinction of structural distortion, whereas their $L_d$ images clearly show the distinguishable, remarkable distributions. As seen in these images, the mass density distribution in all three metastatic cell lines can be clearly differentiated from the mass density distribution observed for normal PWR prostate cells, suggesting greater tumorigenicity of cancer cells based on the degree of disease hierarchy.

In further analyses, 8-12 cells were randomly picked from the micrograph of each cell category to study cell nuclei whose typical size was around 7-10 μm. The structural disorder strengths, $L_d$, were calculated at different length scales (1.25 – 3.5 μm) for each cell categories. The plots are shown in Figure 3(e). The $L_d$ values for each of the prostate cancer cell lines, turned out to be higher than that from the normal PWR prostate cells at all the length scales. The mass density fluctuation observed for AR-independent C4-2 cell lines (Fig. 3(d')) was higher than that calculated for LNCaP cells (Fig. 3(b')). Since C4-2 is derived from cell line LNCaP, it is expected to have higher tumorigenicity and metastatic potential compared to LNCaP, which is in agreement with our results. Figure 3(f) shows bar graphs comparing the calculated disorder strengths for all prostate cell lines examined at the length scale 3.5 μm. The disorder strengths, i.e., mass density fluctuation, calculated for PWR, LNCaP, DU145, and C4-2 cell lines, are in increasing order. The hierarchy of mass density fluctuations observed in the aforementioned prostate cells also agrees with their known tumorigenicity level. Student's $t$-test shows p-value < 0.05 for each pair of the statistical data, suggesting that the mean disorder strengths calculated for the different cell lines are significantly different.

The increase in disorder strength $L_d$ with the increase in tumorigenicity level can be attributed to the changing DNA configuration inside the nucleus. In progressive carcinogenesis, the disorder strength has been linked with the local clumping or condensed configuration forms of the chromatin inside the nucleus[30]. In terms of the parameter studied in this work, i.e. the mass density fluctuation, the present results provide a significant insight as well. For cells with lower tumorigenicity level, the nuclear mass-density corresponding to the DNA led configurations is more uniformly distributed than those for the cells with higher tumorigenicity level. The progression of carcinogenesis proceeds with increase in localized clumping and more randomly arranged configurations of the chromatin structure. Such alteration results in increasing the degree of structural disorder of the cellular structure.

**Correlation between Structural Disorder and Tumorigenicity and hierarchy**: The outcomes of the present study are promising towards quantifying the structural alternations inside the cells. Such a method can be highly useful in diagnosing diseases, e.g. cancer, as well as examining treatment response in therapeutic cancer studies. However, to perform such examination it will require developing robust calibration curves for each separate diagnosis. With this view, we present a framework for a potential calibration curve for analyzing tumorigenicity level in carcinogenesis. A calibration curve for real application purposes should be developed with data from extensive clinical studies.

Figure 4 shows a framework for calibration curve for the prostate cancer case. The structural disorder, as examined for the sample size $3.5 \times 3.5$ μm$^2$ and measured for all four types of prostate cells, has been drawn with respect to the increasing order of tumorigenicity in a single-line $L_d$ spectra. Therefore, each block in Figure 4 represents the range of degree of structural disorder in one type of prostate normal/cancer cell line measured in the present study. Each block is marked in the middle with their mean $L_d$ values 2.8925, 2.9993, 3.0876, and 3.1835, corresponding to the PWR, LNCaP, DU145, and C4-2 cell lines, respectively. This representative calibration line plot, with the quantified degree of structural disorder for a range of prostate cancer cell lines, attempts a quantitative evaluation of tumorigenicity and metastatic potential in any unknown prostate cancer cell type. Such a calibration chart demonstrates the capacity to predict cancer stage and, hence, patient prognosis and survival potential.

*In summary*, the work presented herein describes a novel method to extract mass density (or refractive index) fluctuations information inside the cell nuclei by using confocal fluorescence microscopy in the length scale of ~200 nm to ~2 μm. By analyzing the optical eigenfunction or light localization properties of the cells, we are able to quantify the nano to submicron scale structural disorder in nuclear DNA. This novel technique allows us to characterize the degree of structural disorder using a single parameter, $L_d = dn \times l_c$, which permits a better comparison of the structural alteration in heterogeneous media, in particular, in progressive carcinogenesis. The results of our studies with breast and prostate normal and cancer cells demonstrate that the mass density fluctuation information, or the degree of structural disorder $L_d$, of a biological cell represents a potential parameter to characterize cells. In particular, this information can be used as a potential biological biomarker, numerical index to evaluate stage of cancer cells. One of the important aspects of the present work *is that this method performs quantification of structural alterations in cells irrespective of the factors causing it.* Therefore, such technique can be further extended to study a myriad of cases ranging from various diagnostics to treatment responses at cellular levels. For example, assessing cellular structure in radiation based damage and therapy, examining the effect of targeted drug delivery processes in cell, such as response of anti-cancerous drug, etc. Furthermore, *the method can also be employed for studying non-cancerous cells*, such as red blood cells with sickle cell anemia, stress induced deformities in cells, etc. Overall, our results suggest that this method could add a new dimension to confocal imaging by enabling quantitative characterization of the degree of structural disorder in cells in both health and disease.

**Acknowledgements:**

This work was supported by the NIH grant R01 EB003682 and the University of Memphis to PP, and NIH R01 and U01 grants (UTHSC) to SC. We thank Lauren Thompson for helping us with confocal imaging of breast normal/cancer cells.

**References:**

1. Lodish, H. *et al. Molecular Cell Biology*. (W. H. Freeman, 2000).
2. *Goodman and Gilman's The Pharmacological Basis of Therapeutics, Twelfth Edition*. (McGraw-Hill Education / Medical, 2011).
3. *Principles and Methods of Toxicology, Fifth Edition*. (CRC Press, 2007).
4. Bernstein, C., Bernstein, H., Payne, C. M. & Garewal, H. DNA repair/pro-apoptotic dual-role proteins in five major DNA repair pathways: fail-safe protection against carcinogenesis. *Mutat. Res.* **511,** 145–178 (2002).
5. Vilenchik, M. M. & Knudson, A. G. Endogenous DNA double-strand breaks: production, fidelity of repair, and induction of cancer. *Proc. Natl. Acad. Sci. U. S. A.* **100,** 12871–12876 (2003).

6. Ames, B. N., Shigenaga, M. K. & Hagen, T. M. Oxidants, antioxidants, and the degenerative diseases of aging. *Proc. Natl. Acad. Sci.* **90,** 7915–7922 (1993).

7. Subramanian, H. *et al.* Optical methodology for detecting histologically unapparent nanoscale consequences of genetic alterations in biological cells. *Proc. Natl. Acad. Sci.* **105,** 20118–20123 (2008).

8. Pradhan, P. *et al.* Quantification of nanoscale density fluctuations using electron microscopy: Light-localization properties of biological cells. *Appl. Phys. Lett.* **97,** 243704 (2010).

9. BMC Biophysics | Full text | Investigation of nanoscale structural alterations of cell nucleus as an early sign of cancer. at <http://www.biomedcentral.com/2046-1682/7/1>

10. Schmitt, J. M. & Kumar, G. Optical Scattering Properties of Soft Tissue: A Discrete Particle Model. *Appl. Opt.* **37,** 2788 (1998).

11. Schmitt, J. M. & Kumar, G. Turbulent nature of refractive-index variations in biological tissue. *Opt. Lett.* **21,** 1310 (1996).

12. Pradhan, P. & Sridhar, S. Correlations due to Localization in Quantum Eigenfunctions of Disordered Microwave Cavities. *Phys. Rev. Lett.* **85,** 2360–2363 (2000).

13. Pradhan, P. & Sridhar, S. From chaos to disorder: Statistics of the eigenfunctions of microwave cavities. *Pramana* **58,** 333–341 (2002).

14. Lee, P. A. & Ramakrishnan, T. V. Disordered electronic systems. *Rev. Mod. Phys.* **57,** 287–337 (1985).

15. Muller, M. *Introduction to Confocal Fluorescence Microscopy*. (SPIE Press, 2006).

16. *Handbook Of Biological Confocal Microscopy*. (Springer US, 2006). at <http://link.springer.com/10.1007/978-0-387-45524-2>

17. Waters, J. C. Accuracy and precision in quantitative fluorescence microscopy. *J. Cell Biol.* **185,** 1135–1148 (2009).

18. Roukos, V., Pegoraro, G., Voss, T. C. & Misteli, T. Cell cycle staging of individual cells by fluorescence microscopy. *Nat. Protoc.* **10,** 334–348 (2015).

19. Darzynkiewicz, Z. Critical Aspects in Analysis of Cellular DNA Content. *Curr. Protoc. Cytom. Editor. Board J Paul Robinson Manag. Ed. Al* **CHAPTER,** Unit7.2 (2010).

20. Swedlow, J. R., Hu, K., Andrews, P. D., Roos, D. S. & Murray, J. M. Measuring tubulin content in Toxoplasma gondii: A comparison of laser-scanning confocal and wide-field fluorescence microscopy. *Proc. Natl. Acad. Sci.* **99,** 2014–2019 (2002).

21. Rigaut, J. P. *et al.* Three-dimensional DNA image cytometry by confocal scanning laser microscopy in thick tissue blocks. *Cytometry* **12,** 511–524 (1991).

22. Visscher, K., Brakenhoff, G. J. & Visser, T. D. Fluorescence saturation in confocal microscopy. *J. Microsc.* **175,** 162–165 (1994).

23. Davies, H. G. & Wilkins, M. H. F. Interference microscopy and mass determination. *Nature* **169,** 541 (1952).

24. Barer, R., Ross, K. F. A. & Tkaczyk, S. Refractometry of Living Cells. *Nature* **171,** 720–724 (1953).

25. Molinari, D. & Fratalocchi, A. Route to strong localization of light: the role of disorder. *Opt. Express* **20,** 18156 (2012).

26. John, S. in *Photonic Band Gap Materials* (ed. Soukoulis, C. M.) 563–665 (Springer Netherlands, 1996). at <http://link.springer.com/chapter/10.1007/978-94-009-1665-4_37>

27. Wiersma, D. S., Bartolini, P., Lagendijk, A. & Righini, R. Localization of light in a disordered medium. *Nature* **390,** 671–673 (1997).

28. Prigodin, V. N. & Altshuler, B. L. Long-Range Spatial Correlations of Eigenfunctions in Quantum Disordered Systems. *Phys. Rev. Lett.* **80,** 1944–1947 (1998).

29. Schwartz, T., Bartal, G., Fishman, S. & Segev, M. Transport and Anderson localization in disordered two-dimensional photonic lattices. *Nature* **446,** 52–55 (2007).

30. The influence of chromosome density variations on the increase in nuclear disorder strength in carcinogenesis - IOPscience. at <http://iopscience.iop.org/article/10.1088/1478-3975/8/1/015004/meta>

**Figure Captions**

**Fig. 1**. Schematic: Construction of a 2D optical grid. Point-to-point mapping from the confocal micrograph to construct the refractive index optical lattice system. (a) confocal stack derived from z-scannig (b); (c) a typical confocal micrograph taken from the middle of the z-stack; (d) Intensity fluctuation in confocal along a line; (e) a typical intensity distribution for a 4 × 4 pixel size and corresponding refractive index matrix (f).

**Fig. 2**. Structural disorder analysis of breast cell lines MCF-10A (normal) and MCF-7 (carcinoma). (a), (b): Representative confocal images of DAPI- stained nuclei from MCF-10A and MCF-7 cells, and (a'), (b') their corresponding $L_d$ images. (c): Structural disorder $L_d$ vs. sample length L. (d): Bar graph comparing the structural disorder level at the sample length L=1.6 µm.

**Fig. 3**. Structural disorder analysis of prostate cells. (a) - (d): Confocal images of a bunch of DAPI-stained nuclei from PWR, LNCaP, DU145, and C4-2 prostate cell lines taken in single shot, and (a') - (d') their corresponding disorder strength $L_d$ images. (e) $L_d$ vs. length scale (sample size) observed for all cells studied. (f) Bar graph comparison at 3.5 µm sample length (sample size).

**Fig. 4**. A proposed representative calibration curve, i.e., structural disorder $L_d$ vs. tumorigenicity level inside the cells, as determined at the sample length of 3.5 µm.

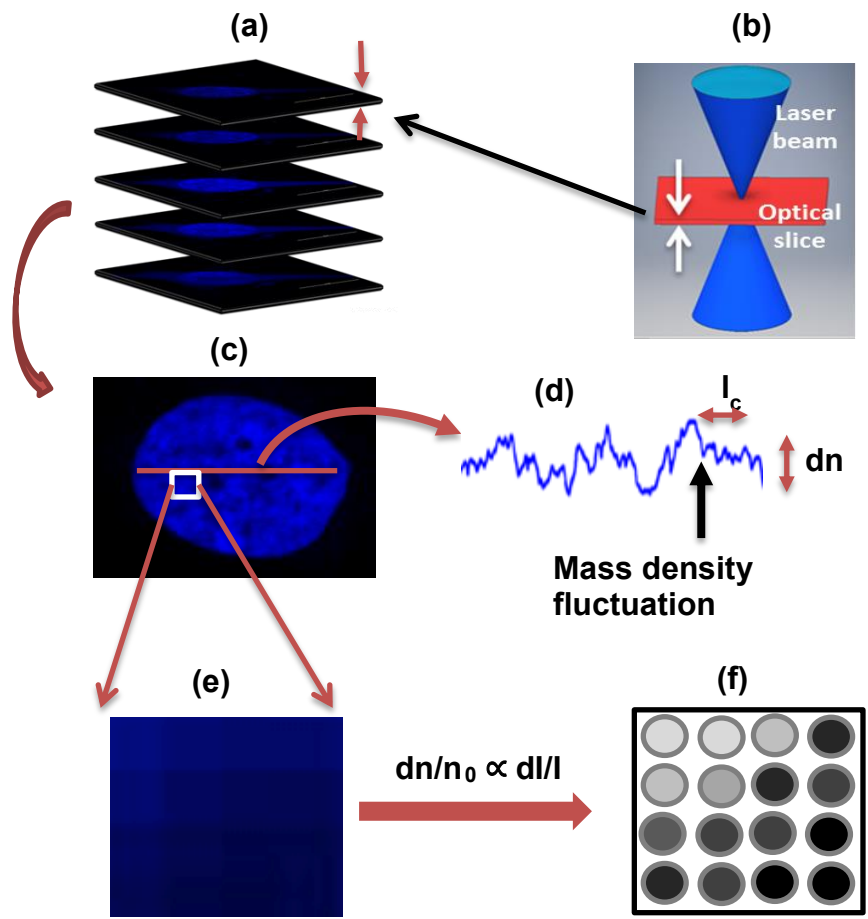

**Fig. 1.**

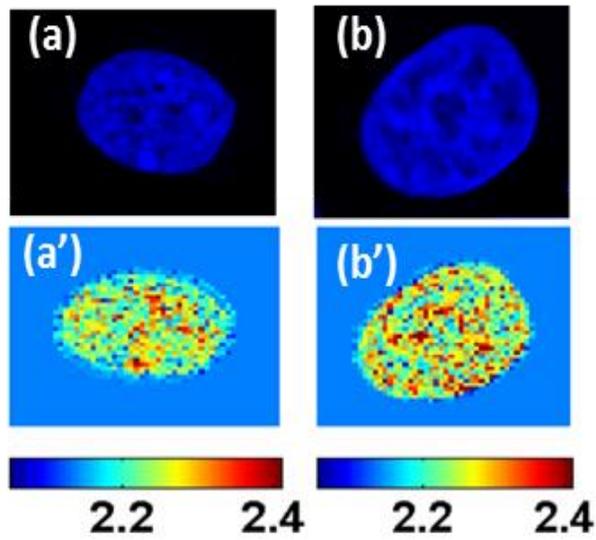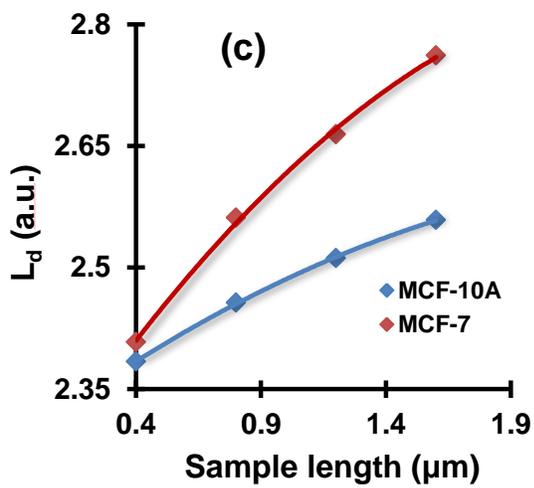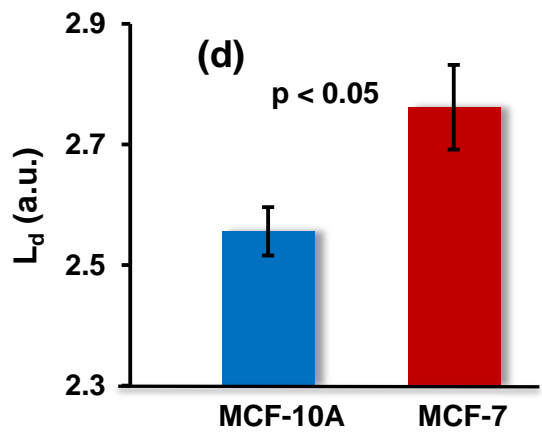

Fig. 2.

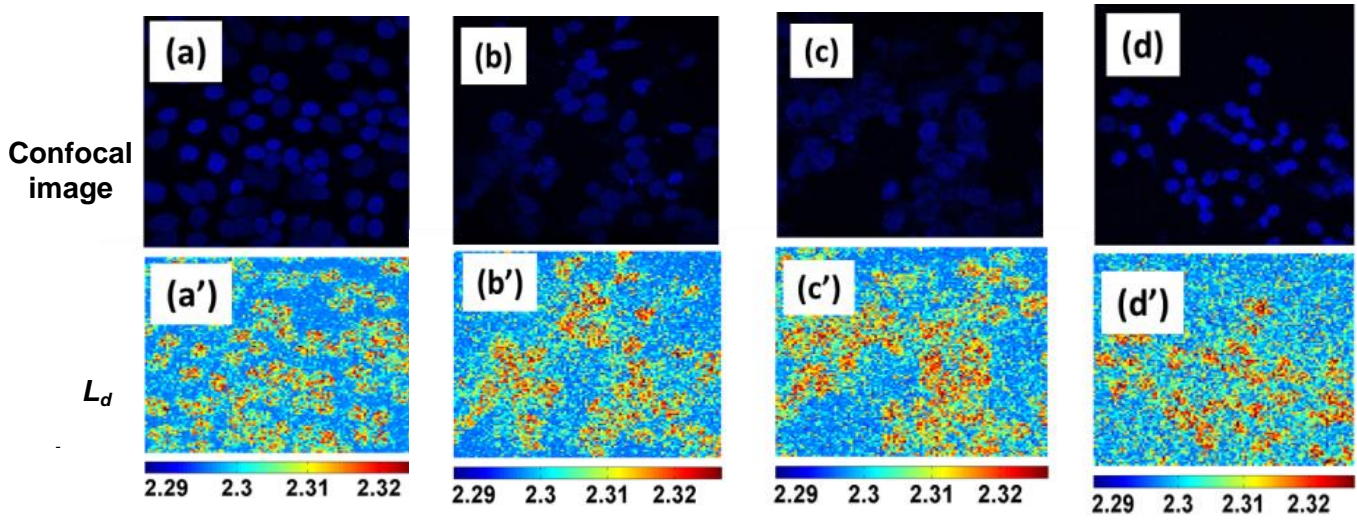

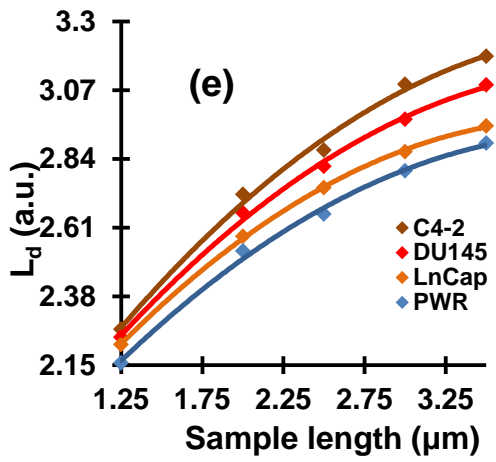
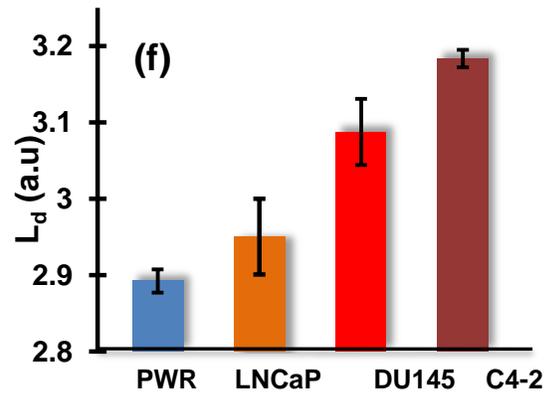

Fig. 3.

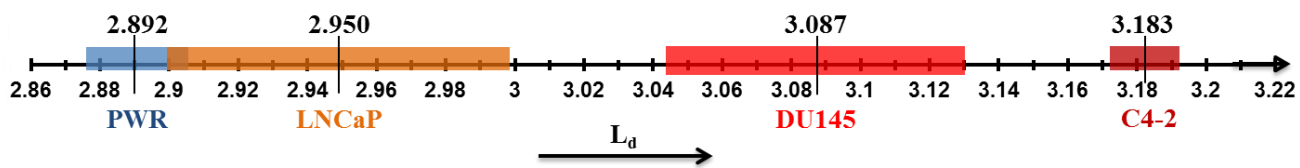

**Fig. 4.**